\documentclass{aa}     
\usepackage[dvips]{graphics}
\usepackage{latexsym}
\begin{document}

\newcommand{\zs} {$\zeta_\star$}
\newcommand{\vsini} {$v\,\sin i$}
\newcommand{\Sc} {Sect.}
\newcommand{\cmc} {cm$^{-3}$}
\newcommand{\cmq} {cm$^{-2}$}
\newcommand{\csqg} {cm$^2$ g$^{-1}$}
\newcommand{\kms} {km s$^{-1}$}
\newcommand{\myr} {$M_\odot$ yr$^{-1}$}
\newcommand{\Myr} {$M_\odot$ yr$^{-1}$}
\newcommand{\um} {$\mu$m}
\newcommand{\mic} {$\mu$m}
\newcommand{\sbr} { erg cm$^{-2}$ s$^{-1}$ sr$^{-1}$}
\newcommand{\sbu} { erg cm$^{-2}$ s$^{-1}$ sr$^{-1}$}
\newcommand{\Lsun} {$L_\odot$}
\newcommand{\Msun} {$M_\odot$}
\newcommand{\Teff} {$T_\star$}
\newcommand{\Tstar} {$T_\star$}
\newcommand{\Lstar} {$L_\star$}
\newcommand{\Rstar} {$R_\star$}
\newcommand{\Mstar} {$M_\star$}
\newcommand{\Md} {$M_{\rm D}$}
\newcommand{\Rd} {$R_{\rm D}$}
\newcommand{\pms} {pre-main--sequence}
\newcommand{\DV} {$\Delta V$}
\newcommand{\Dm} {$\Delta m$}
\newcommand{{\1}} {Cha H$\alpha$1}
\newcommand{{\2}} {Cha H$\alpha$2}
\newcommand{{\9}} {Cha H$\alpha$9}
\newcommand{{\AI}} {$A_{\rm I}$}
 
\newcommand{\simless}{\mathbin{\lower 3pt\hbox
      {$\rlap{\raise 5pt\hbox{$\char'074$}}\mathchar"7218$}}} 
\newcommand{\simgreat}{\mathbin{\lower 3pt\hbox
     {$\rlap{\raise 5pt\hbox{$\char'076$}}\mathchar"7218$}}} 

\title{Exploring Brown Dwarf Disks}

\author {Antonella Natta\inst{1} and Leonardo Testi\inst{1}}
\institute{
    Osservatorio Astrofisico di Arcetri, Largo E.Fermi 5,
    I-50125 Firenze, Italy
}

\offprints{natta@arcetri.astro.it}
\date{Received ...; accepted ...}
 
\titlerunning{BD disks}
\authorrunning {Natta \& Testi}

\begin{abstract}
{We discuss  the spectral energy distribution of three very low mass 
objects in Chamaeleon I for which ground-based spectroscopy and photometry
as well as ISO measurements in the mid-infrared are available
(Comer\'on et al. 2000; Persi et al. 2000). One of these stars
(\1) is a bona-fide brown dwarf, with mass 0.04--0.05 \Msun.
We show that  the observed emission is very well described
by models of circumstellar disks identical to those associated to T Tauri stars,
scaled down to keep the ratio of the disk-to-star mass constant
and to the appropriate stellar parameters.
This result provides a first indication that the formation mechanism
of T Tauri stars (via core contraction and formation of an
accretion disk)
extends to objects in the brown dwarf mass range.
}
\end{abstract}

\maketitle

\section{Introduction}
The last few years have seen an enormous progress in our understanding of sub-stellar mass objects, as
more and more objects in the mass range of brown dwarfs (BD; \Mstar$\simless 0.75$ \Msun) and
giant planets ($\simless$ 0.015 \Msun)
are found, both  in the field and in regions
of recent star formation (Basri~\cite{B00}; Lucas \& Roche~\cite{LR00};
Zapatero Osorio et al.~\cite{ZOea00}).  It is now clear that
``free-floating" very low mass objects exist. Do they form, as
low-mass stars do, from the collapse of a molecular core? 
This, although apparently very reasonable, is not the only
possibility. Alternative hypotheses have been suggested; for example
that BD form 
in gravitationally unstable regions
of protostellar disks (Pickett et al.~\cite {Pk00}) or that they are
stellar embryos, whose further growth is prevented by dynamical 
ejections from small stellar systems (Reipurth \& Clarke~\cite{RC01}).

A full understanding of the formation mechanism(s) of sub-stellar
objects will take  time.
At present, it is timely to begin
to explore the
properties of BD systems 
in regions of star formation and compare them
to the much better
known  \pms\ low-mass stars, the T Tauri stars (TTS).
The most important clue to a formation mechanism
involving accretion from a parental core is the presence of
a circumstellar disk. Claims of evidence of excess emission in
the near (Oasa et al. ~\cite{OTS99}; Muench et al. ~\cite{MLAL01}) 
and mid-IR (Comer\'on et al. ~\cite{CNK00}) in BD 
or BD-candidates in star-forming
regions are beginning to  appear in
the literature. H$_\alpha$  (Comer\'on et al. ~\cite{CRN99};
Comer\'on et al. ~\cite{CNK00}) and
X-ray  emission (Neuhauser \& Comer\'on ~\cite{NC98}; Comer\'on et al. ~\cite{CNK00})
 are seen in some 
very low mass objects.
Muzerolle et al.~(\cite{Mea00}) detect evidence of magnetospheric accretion
at a very low rate $\sim 5\times 10^{-12}$ \Myr\ in the spectrum of
the M6 object
V410 Anon 13, whose mass is estimated in the range 0.04--0.06 \Msun.

In  very cold BD, the most convincing evidence of excess
emission, the accepted signature of circumstellar
disks, can only be obtained in the mid-IR. There are at
present  very few bona-fide BD (i.e., stars with spectroscopical
classification) with measured mid-IR excess. Comer\'on et al.~(\cite{CNK00})
list a small group of BD detected in the mid-IR survey of Cham I
at 6.7 and 14.3 \um\
by ISO (Persi et al.~\cite{Pea00}). Of these, only three have detections
in both bands, the one with the lowest mass being \1, a M7.5 star
with mass  0.04--0.05 \Msun.

In this paper, we show that disk models analogous to those developed
for TTS can well account for the observed spectral energy distribution (SED) of these three stars, providing strong support to the
idea that BD form like the more massive TTS. We will also
speculate somewhat on the derived disk properties, and on the
possibility of extending this kind
of study to objects of  lower mass.

\section {Disk Models}
We have computed the emission expected from a circumstellar disk
heated by the irradiation of the central star following the
method outlined by Chiang \& Goldreich~(\cite{CG97}; CG97), which has
been successfully applied to \pms\ TTS and Herbig Ae stars 
(Natta et al.~\cite{Nea00}; \cite{Nea01}; Chiang et al.~\cite{Cea01}).
CG97 consider flared disks, in hydrostatic
equilibrium in the vertical direction, and make a number of
simplifying assumptions which permit the computation
of  the resulting SED
in a  quick and efficient fashion. Although not entirely self-consistent,
such models provide a good first approximation to the SED, more
than sufficient for the purpose of this paper.

We have taken for the various model parameters values
typical of \pms\ stars, scaled down where necessary.
We have assumed that the circumstellar disk extends inwards
to \Rstar, has outer radius \Rd = 100 AU, total mass
\Md=0.03 \Mstar (Natta et al.~\cite{NGM00}), power-law surface density $\Sigma\propto R^{-1.5}$;
the dust in the disk midplane has opacity 
$\kappa = 0.01 (1.3{\rm mm}/\lambda)$ cm$^2$ g$^{-1}$ (Beckwith et
al.~\cite{Bea90}). 
On the disk surface, we assume the mixture of carbonaceous and silicate grains that provides a good fit to the mid-IR emission of
\pms\ stars  (see Natta et al.~\cite{Nea01} for details).
Most of these parameters are
either irrelevant for the determination of the
mid-IR flux, or appear in combinations,
and cannot be individually constrained by the data available
(see, for a discussion, Chiang et al.~\cite{Cea01}). 
At this stage, only the  most ``standard" assumptions
are justified.

The stellar properties have been determined by 
Comer\'on et al.~(\cite{CNK00}). The most important parameter
for the disk SED calculation is the stellar luminosity and, 
to a lesser degree, the  ratio \Mstar/\Rstar, that controls
the disk flaring angle (roughly $\propto$(\Rstar/\Mstar)$^{4/7}$; CG97). 
We have used in displaying the results of our calculations
the  model stellar atmospheres
of Allard et al.~(\cite{Aea00}; \cite{Aea01}).
When comparing them with the broad-band visual and
near-infrared photometry, we found good agreement only for values of the effective temperatures   significantly lower than
the spectroscopically determined   values of Comer\'on et al.~(\cite{CNK00}).
We do not know if this effect
has any significance, given the uncertainties on the temperature
scale in this mass range.
In any case, our effective temperatures
are well within the range of values expected for field brown dwarfs of
similar spectral type (see, for example, Leggett et al.~\cite{Lea01}), and their
exact value is unimportant for the disk SED determination.
The extinction and 
luminosity we derive   are  identical to the Comer\'on
et al. values for \1\ 
and \2,
while  they are definitely
higher for \9, for which we estimate  a luminosity
of 0.018 \Lsun\  as compared with
Comer\'on et al. value of 0.0056 \Lsun. 
The adopted values of the stellar parameters are given in the figure
captions. We have not re-determined the stellar masses, for which
we  used the Comer\'on et al. estimates.

\begin{figure}
\resizebox{\hsize}{!}{\includegraphics{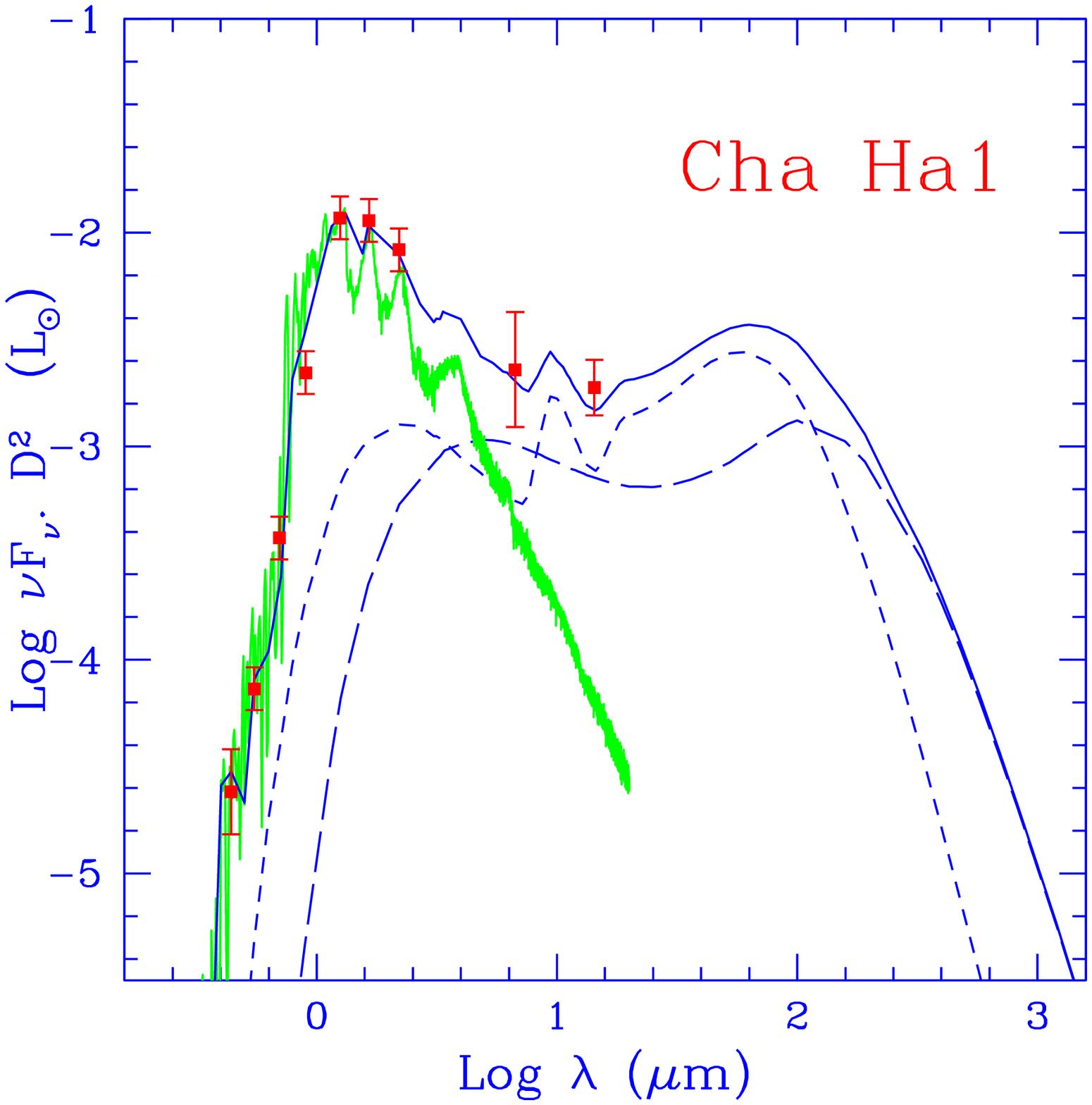}}
\caption{ SED of \1. The star has \Tstar=2400 K, \Lstar=0.01 \Lsun,
\Mstar=0.05 \Msun, extinction in the I band \AI=0.11 mag.
Disk parameters as in the text, the viewing angle is 0$^o$
(face-on).
The jagged line shows the Allard et al. (2001) model atmosphere,
the solid line the disk model predictions. The short and long dashed lines show
separately the contribution to the SED of the disk surface and midplane.
The filled squares with error bars
 are the observed fluxes from Comer\'on et al. (2000).
 }
\label{cha1}
\end{figure}

The most interesting of the three stars is \1\, which has the
lowest mass of the three (about 0.04--0.05 \Msun). The results are shown in Fig.~\ref{cha1}.
The disk predictions fit extremely well the observed points
at all available wavelengths.

\begin{figure}
\resizebox{\hsize}{!}{\includegraphics{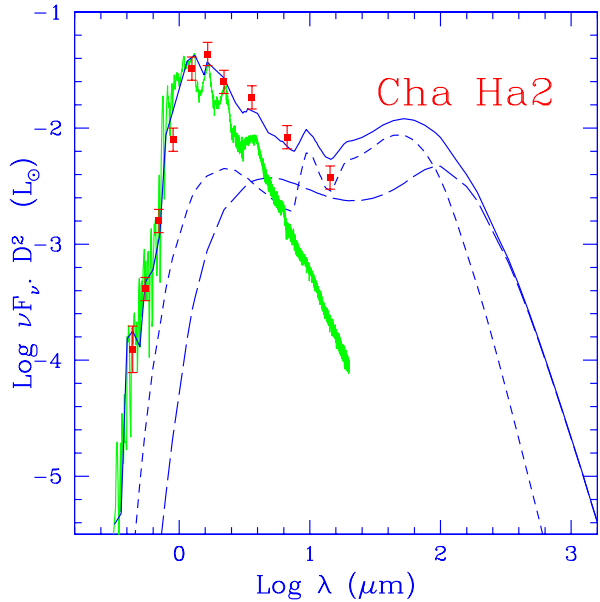}}
\caption{Same as Fig.~1 for \2. The star has \Tstar=2550 K, \Lstar=0.035 \Lsun, 
\Mstar=0.08 \Msun, \AI=0.4 mag. The disk viewing angle is 0$^o$.
Note that the formal errors on the mid-infrared fluxes quoted by Comer\'on et al.
are 6\% and 5\% at 6.7 and 14.3 \um, respectively, while we have plotted
error bars corresponding to a more realistic 10\% uncertainties.
The point at 3.6 \um\ is from Kenyon \& G\'omez 2000.
}
\label{cha2}
\end{figure}

\begin{figure}
\resizebox{\hsize}{!}{\includegraphics{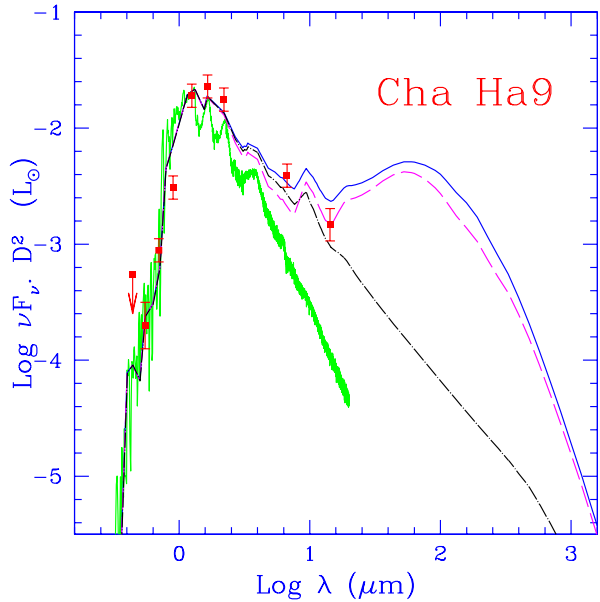}}
\caption{Same as Fig.~1 for \9. The star has \Tstar=2500 K, 
\Lstar=0.018 \Lsun, \Mstar=0.09 \Msun, \AI=1.5 mag. The disk viewing angle
(solid line) is 0$^o$. Also shown are the predictions of the same
disk seen at an inclination of 75$^o$ (dashed line) and of
a flat disk seen at 0$^o$ (dot-dashed line).
To avoid confusion, we do not show the disk surface and midplane
constributions separately. }
\label{cha9}
\end{figure}

The results for the other two stars are shown in Fig.~\ref{cha2}
and \ref{cha9}, respectively. 
Our standard disk model fits rather well the observed photometric points,
although we note a tendency of the 14.3 \um\ point to lie slightly
below the model predictions. This discrepancy, however, is only 
20\% for \2\ and 25\% (2$\sigma$) for \9. For this star, we show in Fig.~3
the predictions of the same disk model seen with an inclination
angle of 
 75 deg (dashed line). The flux is only slightly reduced, since it is dominated at most wavelengths by emission of optically thin material
(see Fig.~\ref{cha1} and \ref{cha2}).
The dot-dashed line shows the SED predicted by
 a geometrically flat disk
seen face-on (dot-dashed line). While the difference at long wavelengths is very large,
the mid-infrared flux
would   still be consistent with the
ISO measurements.
One point worth to notice is that in these low luminosity objects the mir-IR
is emitted by the inner disk, and the ISO points would not be
consistent with a disk inner hole larger than about 3~\Rstar.
 

The results shown in Fig.~1, 2 and 3 provide  good evidence that
the same kind of disks which reproduce the
properties of \pms\ stars exist around lower-mass
objects, including  a bona-fide BD such as \1.
In all  three cases, 
the disks need to be
optically thick in the mid-infrared. This, however, sets only a 
weak constraint on the  mass of the disk, which remains optically thick
at 14.3 \um\ as long as
\Md$\simgreat 1.5\times 10^{-6} \kappa_{14.3\mu m}$ \Msun,
where $\kappa_{14.3\mu m}$ is in units of cm$^2$ g$^{-1}$.
For $\kappa_{14.3\mu m}\sim 3-10$ (Henning \& Stognienko~\cite{HS96}),
this limit translates into \Md$\simgreat 10^{-5}$ \Msun, or
$\simgreat 2\times 10^{-4}$ \Mstar, about ten times lower than
the smallest measured ratios in \pms\ stars 
 (Natta et al.~\cite{NGM00}).
The disk mass, and, to some extent, its size, 
can be better determined
from
millimeter data.
For a typical value \Md$\sim$0.03 \Mstar, we predict
for \1\ a 1.3mm flux of 3 mJy, well within the range of existing
millimeter telescopes.

\1\ has spectral type M7.5 and mass of 0.04--0.05 \Msun. Of course,
one would like to extend our knowledge of circumstellar disks
to even lower mass objects,  such as are currently being discovered
in the Orion Nebula Cluster (ONC; Lucas \& Roche~\cite{LR00})
and in $\sigma$ Orionis (Zapatero Osorio et al.~\cite{ZOea00}).
Fig.~4 shows the predicted SED for objects of lower and lower mass, 
as labelled, assuming an age of 2 Myr. Photospheric
effective temperatures and
luminosities are from Baraffe et al.~(\cite{Bea98}) and Burrows et
al.~(\cite{Bea01}). 
Given the heuristic purpose of this figure,
we
have assumed  that the stellar emission can be
described by a black body
 at \Tstar. This, as already noted, does not affect
the calculations of the disk emission, but is a  poor description of the
photospheric spectrum. As a consequence, the SEDs in Fig.~4 are not
realistic at wavelengths shorter
than  $\sim$4 \um, where the photospheric emission dominates.

Our calculations show that in nearby star forming regions it will be possible
to detect disks around young sub-stellar objects with current
instrumentation. Modern mid-infrared cameras at large telescopes have 
a
10~$\mu$m sensitivity allowing detection of  disk emission from systems more
massive than $\sim$10~M$_J$. The upcoming space missions 
(SIRTF, HERSCHEL and especially NGST)
will allow 
to detect  disk emission around  planetary-mass objects in
the ONC and the $\sigma$~Orionis clusters. Dust spectroscopy in the
mid-infrared, as obtained by ISO for luminous \pms\ A stars (see
Waelkens et al. ~\cite{WISO96}),
is beyond the capability of even the largest ground-based telescopes,
but for the most massive and nearby BD;
in principle SIRTF and NGST will allow the observation
of the  spectra
of lower mass or more distant objects, although the former satellite may
be seriously affected by confusion problems.
Perhaps the most interesting
observations,   determining the amount of circumstellar
material around these objects and possibly  probing
 the disk kinematics, will
be those in the millimeter-wave range. As noted above,
our disk model for
Cha~H$\alpha$1 predicts a continuum flux (about 3 mJy at 1.3 mm)
well within the range of
detectability with current instrumentation. To detect disks around less massive
and more distant systems and to attempt the detection of molecular line
emission in order 
to study the disk kinematics, we will need to wait for the ALMA
array to be operational.

\begin{figure}
\resizebox{\hsize}{!}{\includegraphics{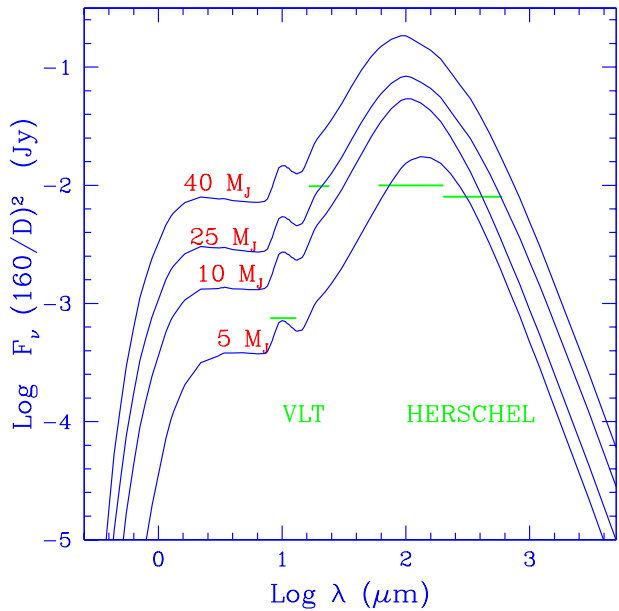}}
\caption{ Plot of the expected flux for
BD of decreasing mass, from 40 Jupiter mass (0.04 \Msun)
to 5 M$_J$, as labelled.  Below $\sim$4 \um, the emission is dominated by
the stellar photosphere, assumed here as a black body at \Tstar.
The expected detection limit (10$\sigma$ in 1 hr) are shown for
the VLT/VISIR mid-infrared camera and for the far-infrared
bolometers on HERSCHEL (horizontal bars).}
\label{many}
\end{figure}

\section {Conclusions}

The presence and properties of disks around very low mass objects,
which are  being
discovered in regions of star formation,  are of crucial importance for
understanding  their formation.
If disks exist, and if their properties are analogous to those
of disks around TTS, we can  conclude that the formation mechanism
of TTS, by collapse and accretion of a molecular core, is very likely
to extend
to the lowest mass objects.
 
This paper presents the first attempt to test quantitatively
this hypothesis.
We have modeled the emission expected from disk models of properties
identical to those of TTS for three objects in Chamaeleon I. These
objects have been studied spectroscopically by Comer\'on et al. (2000),
who attributed to them spectral types M7.5--M6 and masses in the range
0.04--0.09 \Msun. The lowest mass object (\1) is certainly a bona-fide
BD. These objects are
 of particular importance because they have been detected
at 6.7 and 14.3 \um\ by ISO (Persi et al. 2000), providing a unique
(until now) possibility to constrain the disk properties of BD.
Inspection of Fig.1 to 3 shows that it is extremely difficult 
to infer the existence of  disks from  near-infrared photometry,
where the emission is largely dominated by the stellar photosphere.

In all  three cases,
and in particular in the most interesting object  \1,
optically thick, flared disks  are required
to account for the mid-infrared fluxes. The disks are heated by the
central star, and we expect that relatively small grains on the disk
surface contribute significantly to the observed mid-infrared
emission. If higher spectral resolution data in this range were
available, we predict that one would
observe the 10 \um\ silicate feature in emission.
Such disks are identical to those around TTS, just scaled to 
the appropriate stellar parameters.

This result seems to us very interesting. It provides a first
indication that
the TTS formation mechanism (from a collapsing core
via an accretion disk) extends to objects of 0.04--0.05 \Msun.
It does not, however, rule out the possibility that BD
are ejected stellar embryos, as suggested by Reipurth \& Clarke (2001).
In their model, the embryos may keep a small circumstellar disk,
of few AU size. The existing data, limited to wavelengths shorter than
$\sim 15$ \um, can only set a limit to the disk radius of \Rd$\simgreat$1 AU.
There are two ways to proceed, the first and more obvious
is to detect the millimeter emission of at least some BD,
which will be well below detection in the embryo hypothesis.
The second is to 
search
for other BD and planetary mass objects
in the  mid-infrared.
A high frequency of disks is not predicted by
the Reipurth \& Clarke model, since the small, truncated disks,
associated with the embryos, not
fed by any surrounding core, will rapidly disappear.

\begin{acknowledgements}
This work was partly supported by  ASI grant  ARS 1/R/27/00 to the
Osservatorio di Arcetri.
\end{acknowledgements}

\end{document}